\documentclass{ws-procs9x6}

\begin{document}

\title{Charmonium correlators at finite temperature
in quenched lattice QCD%
\footnote{\uppercase{T}alk presented by
     \uppercase{H}.~\uppercase{M}atsufuru.}}

\author{Hideo Matsufuru$^{\rm a}$, Takashi Umeda$^{\rm a}$,
 Kouji Nomura$^{\rm b}$}

\address{
 $^{\rm a}$Yukawa Institute for Theoretical Physics, Kyoto University,\\
   Kyoto 606-8502, Japan \vspace{-0.4cm}}

\address{
 $^{\rm b}$Department of Physics, Hiroshima University,\\
   Higashi-hiroshima 739-8526, Japan}

\maketitle

\abstracts{
We study charmonium correlators at finite temperature
using quenched lattice QCD simulations.
Two analysis procedures are applied to extract information
on the spectral function: the maximum entropy method, and
the $\chi^2$ fit analyses including the constrained curve fitting.
We focus on the low energy structure of the spectral function
by applying the smearing technique.
We first discuss the applicability of these methods to the finite
temperatures by analyzing the data at $T=0$ with restricted
numbers of degrees of freedom.
Then we apply these methods to the correlators at $T>0$.
We find no indication of mass shift and finite width for the
charmonium states at $T\simeq 0.9T_c$.
The results at $T\simeq 1.1 T_c$ imply that bound-state-like
structures may survive even above $T_c$.}

\section{Introduction}

Investigation of the QCD phase transition at finite temperature and
density is one of the most important subjects in hadron physics.
To create the quark gluon plasma (QGP) phase, in which quarks and
gluons are deconfined and the chiral symmetry is restored, 
the high energy heavy ion collision experiments have been performed
and the RHIC experiment at BNL is in progress.
The $J/\psi$ suppression has been considered as one of the most important
signals of formation of QGP.
However, theoretical understanding of charmonium at finite temperature
and density is not sufficient to draw a conclusive scenario.

We investigate the charmonium properties at finite temperature
using lattice QCD simulations which enable us to incorporate
the nonperturbative effects of QCD.
Observations of spatial correlations between quark and antiquark
suggested nontrivial strong correlations persist above $T_c$
\cite{TARO01,umeda01}.
These imply that bound-state-like structures may survive
even above $T_c$.

In present work, we analyze the correlators by reconstructing
the spectral function from lattice data using the maximum entropy
method \cite{nakahara99} and the $\chi^2$ fit analysis
\cite{ishii02,umeda02}.
The latter includes the constrained curve
fitting \cite{lepage01} in addition to the standard $\chi^2$ fit.
To circumvent severe restriction of degrees of freedom in Euclidean
temporal direction, we employ anisotropic lattices on which
temporal lattice spacing is finer than the spatial one
(for our setup of the anisotropic lattices, see Ref.~\cite{matsufuru01}).
At this stage, the simulations are performed in the quenched
approximation (without dynamical quark effects).
To enhance the contribution from low energy part in the spectral
function, we apply the smearing technique by spatially
extending the meson operators.
The details of this work has been presented in Ref.\cite{umeda02}.

\section{Analysis procedures}

The spectral function contains direct information
on the structure of the mesonic correlator,
$C(t)=\sum_{\vec{x}} \langle O(\vec{x},t) O^{\dag}(0,0) \rangle$.
Recently the maximum entropy method (MEM) has been successfully
applied to the extraction of the spectral function from lattice
data at $T=0$ without assuming specific form \cite{nakahara99}.
However, its applicability to problems at $T>0$ is not straightforward,
because available physical range of $C(t)$ as well as number of
data points are severely limited.
In such a circumstance, we need in some way to check the applicability
of the method.
Once a presumable form of the spectral function is known by MEM,
other analysis procedures, such as the standard $\chi^2$ fit
may provide more quantitative information \cite{ishii02,umeda02}.
We therefore propose to use the $\chi^2$ fit analysis
in combination with MEM \cite{umeda02}.
As a more sophisticated alternative to the standard $\chi^2$ fit,
we also apply the constrained curve fitting \cite{lepage01}
assuming several peaks with finite widths.
In this method, one can add prior knowledge for the spectral function
and thus can treat functional forms with numbers of parameters (even)
more than the number of data points.

To verify the applicability of these methods at $T>0$,
we require that they correctly work for the correlators at $T=0$
with restricted numbers of data points corresponding to the cases
at $T>0$.
As we will see in the next section, MEM applied to the correlators
of the point operators does not meet this condition with present level
of statistics ($O(500)$ configurations).

We therefore apply the smearing technique
which enhances the low energy part of the spectral function.
For the correlators smeared with the wave function at $T=0$,
both the procedures work well in the low energy region.
Although the features of the collective mode such as the mass and
width are unchanged by the smearing,
the smearing can produce a mimic peak structure in the spectral
function \cite{wetzorke02}. 
This possibility is examined by observing the correlators
smeared with a narrower function ({\it half-smeared} correlators).
The dependence of extracted spectral function on the smearing
function is compared with that of correlators composed of free
quarks.

\section{Numerical results}

The numerical simulation is performed on lattices with the spatial
lattice cutoff
$a_\sigma^{-1}\simeq 2$ GeV and the anisotropy $a_\sigma/a_\tau=4$
in the quenched approximation \cite{matsufuru01}.
The lattice sizes are $20^3\times N_t$, where $N_t=160$ ($T\simeq 0$),
$32$ ($T\simeq 0.9 T_c$), and $26$ ($T\simeq 1.1 T_c$).
The numbers of configurations are 500 at $T\simeq 0$ and 1000 
at $T>0$.
$N_t=28$ roughly corresponds to the transition temperature.
The quark field is described by the $O(a)$ improved
Wilson action with the tree-level tadpole-improvement.
The hopping parameter is chosen so that the charmonium spectrum
is roughly reproduced.

\subsection{Results at $T=0$}

\begin{figure}[t]
\centerline{\epsfxsize=2.14in\epsfbox{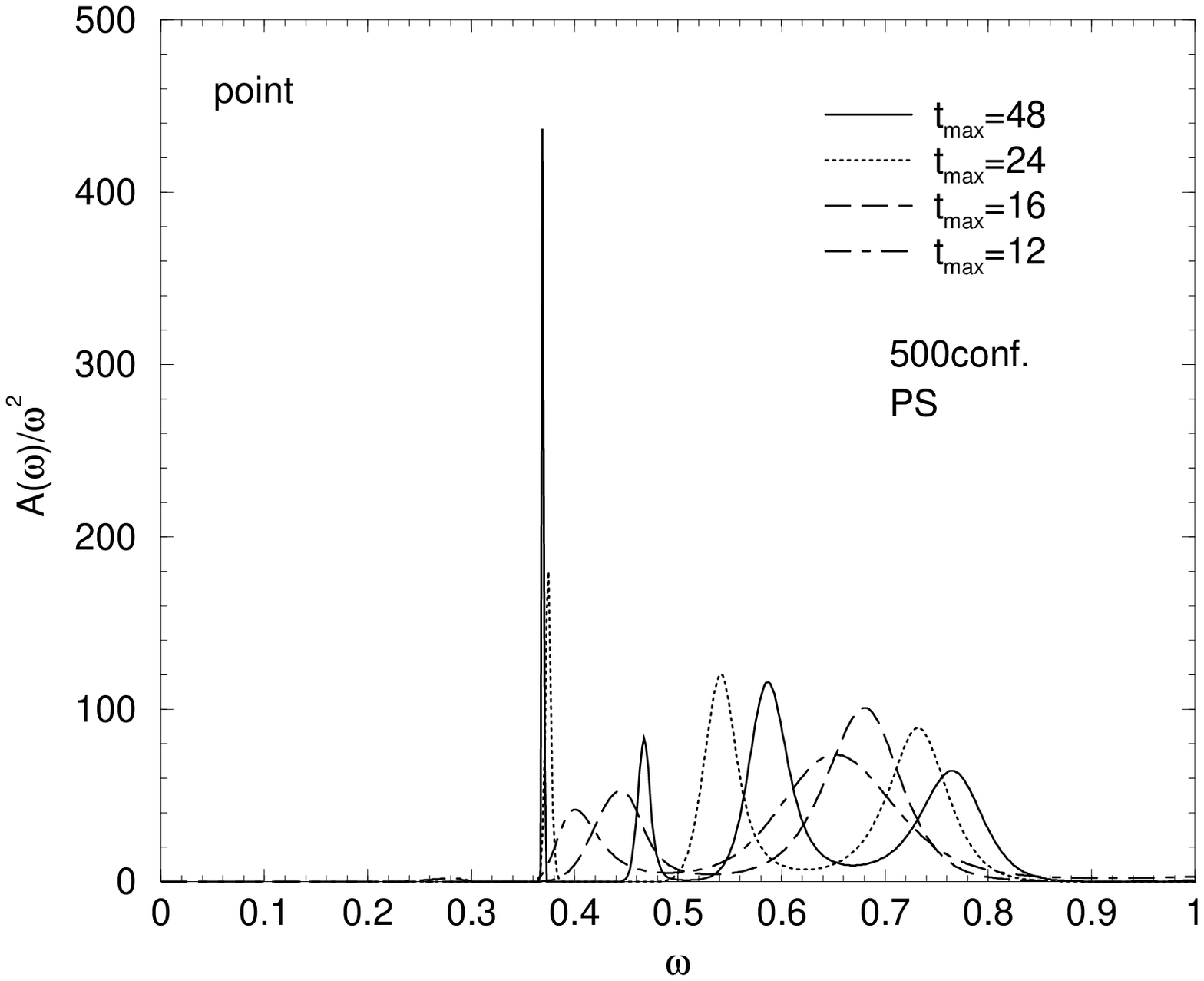}
            \epsfxsize=2.14in\epsfbox{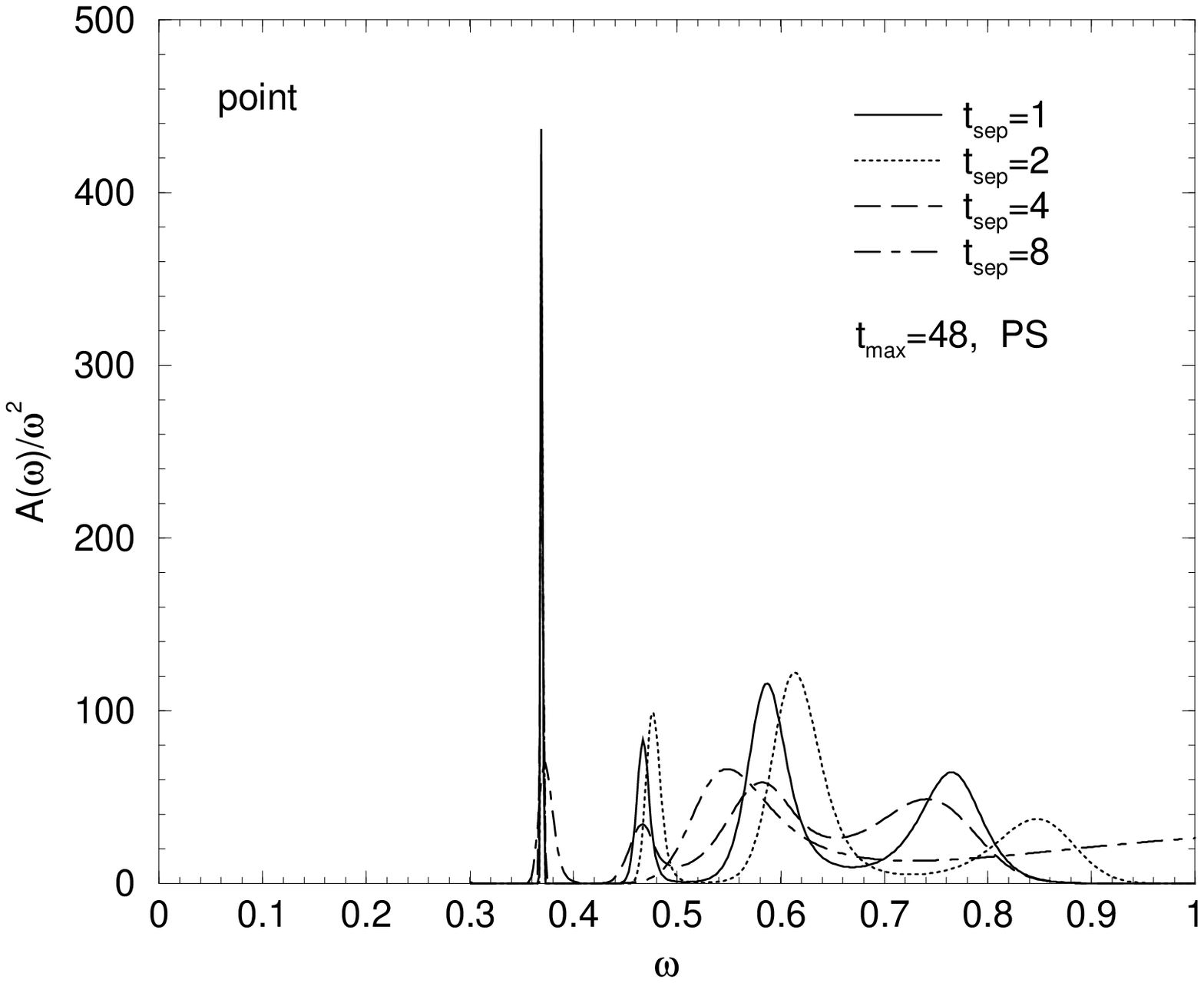}}
\vspace{-0.1cm}
\caption{
The spectral functions determined by MEM for the point correlator
at $T=0$ in PS channel.}
\label{fig:T=0}
\end{figure}

Let us start with the analysis of the point correlators at $T=0$.
We apply MEM to them with restricted numbers of degrees of freedom.
The results with two types of such restrictions are displayed in
Fig.~\ref{fig:T=0}.
The left panel shows the dependence of the result on $t_{max}$,
the maximum $t$ of the correlator used in the analysis.
This case corresponds to the situation at $T>0$.
MEM fails to reproduce even the lowest peak for $t_{max}\leq 16$.
The right panel shows the results when one alternatively skips several
time slices in the analysis.
This case corresponds to the coarsening of the temporal lattice spacing.
Even for $t_{sep}=8$ for which the number of data points is 6,
MEM at least reproduces the correct lowest peak position
while the resolution is not enough.
These results indicate that the physical region of the correlator
as well as the number of the degrees of freedom is important
for MEM to work correctly.
The required region of $C(t)$ in the above analysis is
$t_{max} > O(0.5 \mbox{fm})$, which is not fulfilled around
$T\simeq T_c$.

This is why we apply the smearing technique.
The results of MEM for the smeared correlators are stable under the
above two kinds of restriction for $t_{max}$ of interest;
at least the lowest peak position is correctly reproduced while
the resolution becomes worse as $t_{max}$ decreases.
Since the smearing suppresses the high frequency region of the
spectral function, we focus only on the lowest peak of the spectral
function in the following analysis.

The results of the $\chi^2$ fit and the constrained
curve fitting for the smeared correlators are consistent with
the results of MEM and stable under the same restriction
of $t_{max}$.
In the following, we concentrate on the results of the smeared
correlators.

\subsection{Results at $T>0$}

\begin{figure}[t]
\centerline{\epsfxsize=3.1in\epsfbox{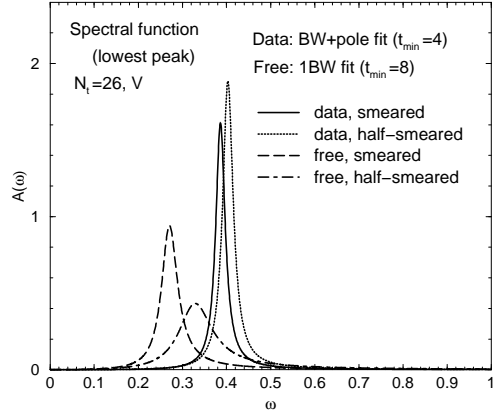}}
\vspace{-0.7cm}
\caption{
The spectral functions from the $\chi^2$ fit analysis
for the smeared correlators in the vector channel at $T\simeq 1.1T_c$.}
\label{fig:SPFs}
\end{figure}

At $T\simeq 0.9T_c$, the result of MEM for the smeared correlator
indicates that the low energy structure of the spectral function is
well represented by a strong peak corresponding to a meson state.
We then apply the $\chi^2$ fit analysis assuming double-pole, single
relativistic Breit-Wigner (BW) type form \cite{ishii02}, and BW+pole
form (the pole function is to subtract the excited state contribution).
The analysis indicates that the lowest peak of the spectral function is
well represented by the pole form with the width consistent with zero.
The mass is almost the same as at $T=0$: no mass shift is observed
at $T\simeq 0.9T_c$.
The result of the constrained curve fitting is consistent with that of
the standard $\chi^2$ fit.

At $T\simeq 1.1T_c$ the result of MEM still exhibits a peak structure 
around the same energy region as at $T<T_c$.
Therefore we perform the same kind of the $\chi^2$ fit analysis as
at $T<T_c$.
The fits to the 2-pole {\it ansatz} and to the Breit-Wigner type
{\it ans\"atze} give
inconsistent results, and the latter fits indicate that the spectral
function has a peak with almost the same mass as at $T<T_c$ and 
the width of order of 200 MeV, as shown in Fig.~\ref{fig:SPFs}.
The result depends on the smearing function only slightly,
in contrast to the free quark case, and hence we conclude that
this is a physically significant structure.
Similar result is obtained with the constrained curve fitting.
These results are in accord with our earlier analysis of the spatial
correlation between quark and antiquark \cite{umeda01}
and imply that quasi-stable bound-state-like structures exist
at $T\simeq 1.1 T_c$.
Similar results have also been reported by other groups
\cite{petretzky03,asakawa03}.

The simulation has been done on
NEC SX-5 at Research Center for Nuclear Physics, Osaka University and
Hitachi SR8000 at KEK (High Energy Accelerator Research Organization).
H.~M. and T.~U. are supported by Japan Society for the Promotion of
Science for Young Scientists.

\end{document}